\documentclass{emulateapj}
\slugcomment{{\sc Accepted to ApJ} March 18th, 2011}

\def\msol{\hbox{$\rm\thinspace M_{\odot}$}}  \def\etal{{\it et al\ }}
\def\eg{{\it e.g.\ }}   
   
   \def\p3m{P${}^3$M}
\def\ap3m{AP${}^3$M} \def\-{{\em{---}}} 
\def\rhobar{{\bar{\rho}}}
\def\svel{{\partial{\widetilde{u}}}}
\def\spos{{\partial{x}}}

\newcommand{\be}{\begin{equation}}  \newcommand{\ba}{\begin{eqnarray}}
\newcommand{\ee}{\end{equation}}  \newcommand{\ea}{\end{eqnarray}}

 \newcommand{\bi}{\begin{itemize}}
\newcommand{\ei}{\end{itemize}}

\def\lesssim{\mathrel{\hbox{\rlap{\hbox{\lower4pt\hbox{$\sim$}}}\hbox{$<$}}}}
\def\gtrsim{\mathrel{\hbox{\rlap{\hbox{\lower4pt\hbox{$\sim$}}}\hbox{$>$}}}}

\usepackage{epstopdf} \usepackage{subfigure} \usepackage{rotating} \usepackage{amsmath}

\begin{document}

\title{Formation of Compact Stellar Clusters by High-Redshift Galaxy Outflows II: Effect of Turbulence and Metal-Line Cooling}
\author{William J. Gray\altaffilmark{1}, \& Evan Scannapieco\altaffilmark{1}}

\altaffiltext{1} {School of Earth and Space Exploration, Arizona State University, P.O. Box 871404, Tempe, AZ, 85287-1494.}

\begin{abstract}

In the primordial universe,  low mass structures with virial temperatures less than 10$^{4}$ K were unable to cool by atomic line transitions, leading to a strong suppression of star formation. On the other hand, these ``minihalos" were highly prone to triggered star formation by interactions from nearby galaxy outflows.    In Gray \& Scannapieco (2010), we explored the impact of nonequilibrium chemistry  on these interactions.  Here we turn our attention to the role of metals, carrying out a series of high-resolution three-dimensional adaptive mesh refinement simulations that include both metal cooling and a subgrid turbulent mixing model.   Despite the presence of an additional coolant, we again we find that outflow-minihalo interactions produce a distribution of dense, massive stellar clusters. We also find that these clusters are evenly enriched with metals to a final abundance of Z $\approx$ 10$^{-2}$ Z$_{\odot}$. As in our previous simulations, all of these properties suggest that these interactions may have given rise to present-day halo globular clusters. 

\end{abstract}


\section{Introduction}

Turbulent processes are instrumental in understanding a wide range of astrophysical observations, including the elemental homogeneity in field stars (Reddy \etal 2003),  open clusters (e.g., Friel \& Boesgaard 1992; Twarog \etal 1997; Carraro \etal 1998), the Magellanic clouds (Olszewsit \etal 1991), dwarf-irregular galaxies (Thuan \etal 1995), and galactic and disk H II regions (e.g., Deharveng \etal 2000, Henry \& Worthey 1999). Turbulent mixing is also important for the enrichment of primordial gas (Pan \etal 2007) and the transition from Population III to Population II stars (Scannapieco \etal 2003). 


Similarly, turbulence affects the distribution and formation of molecular species. Because most chemical reactions are strongly temperature-dependent, by simply moving material to a region with different physical properties such as density, temperature, and UV flux,  or by creating a local heating event through turbulent dissipation,  many reactions can be greatly enhanced, which alters the final abundance of each species (for a review see Scalo \& Elmegreen 2004). This can be particularly important in primordial gas at high-redshift, whose cooling properties are highly dependent on the mass fraction of molecular hydrogen and hydrogen deuteride.  Furthermore, turbulent transport is equally important after reactions occur, as it will alter the physical distribution of the newly formed chemical species (e.g, Xie \etal 1995). 

In the first paper of this series we explored the effect of nonequilibrium chemistry and associated cooling in the interaction of a high-redshift galactic outflow and a low-mass primoridal cloud (Gray \& Scannapieco 2010, hereafter Paper I), in the absence of turbulence or metals.  A large population of high-redshift gravitationally bound clouds with virial temperature below $10^4$ K is a generic prediction of the cold dark matter (CDM) model (Bromm \& Larson 2004 and references within). These clouds, which are often called ``minihalos," are too small to excite the necessary radiative transitions to cool via atomic hydrogen and helium lines. Thus further minihalo cooling requires either molecular-line transitions or metal/dust cooling. Although some small abundance of H$_2$ is expected after recombination and important in cooling of larger structures (e.g., Abel \etal 2002; Bromm \etal 2002) the resulting Lyman-Werner photons (11.2-13.6 eV) from the stars in these objects (e.g., Haiman \etal 1997; Ciardi \etal 2000; Sokasian \etal 2004; O'Shea \& Norman 2007) would likely dissociate this trace amount. Furthermore, even if some trace amount of H$_2$ exists it is unlikely that it would impact the structure of these objects (Whalen \etal 2008a; Ahn \etal 2009).  

In Paper I we used high-resolution three-dimensional adaptive mesh refinement (AMR) simulations to model the interactions between minihalos and high-redshift galactic outflows, including a 14 species model of primordial chemistry.    We found that several dense clusters with masses above 10$^{4}$ M$_{\odot}$ were formed, which may resemble present day halo globular clusters. 

Here we consider the role of turbulent mixing on these interactions, returning to this model and including the effects of unresolved turbulence and metal-line cooling.   In particular, we are interested in three primary questions: First, how does the inclusion of turbulence affect the mixing and final abundance of metals coming from the galactic outflow? Second, does metal-line cooling significantly alter the final state of the cloud? And finally, do we obtain a similar distribution of clusters as our previous simulations without subgrid turbulence or metals?

The structure of this paper is as follows. In Section 2 we describe our approach to the subgrid modeling of turbulence, compare the model to analytic test problems and laboratory experiments, and describe our method for metal line cooling. In Section 3 we outline our model setup and initial conditions, and in Section 4 we present our results and their implications to local observations. Conclusions are given in Section 5.

\section{Numerical Method}

  All simulations were performed using FLASH version 3.1.1, a multidimensional adaptive mesh refinement hydrodynamic code (Fryxell \etal 2000) which solves the Riemann problem on a Cartesian grid using a directionally-split piecewise parabolic method (PPM) solver (Colella \& Woodward 1984; Colella \& Glaz 1985; Fryxell \etal 1990). 

In Paper I, we developed and verified a 14 species chemical network that traced the evolution of both atomic (H and He) and molecular (H$_2$ and HD) species, which included all the pertinent cooling rates. In the current work, in order to study the evolution and impact of metals in outflow-minihalo interactions we add two further packages to our simulations:  a turbulence model that tracks the subgrid mixing of enriched and primordial gas and a cooling module that accounts for additional cooling in the presence of metals.  Here we describe each of these in turn. 

\subsection{K-L Turbulence Model}

	Within FLASH we have implemented a buoyancy and shear driven model of turbulence using a two equation $K$-$L$ model,  where $K$ represents the turbulent kinetic energy and $L$ represents the eddy length scale.  The model was originally developed and used with great success to describe turbulent fluid flow in inertial confinement fusion (ICF) experiments (Dimonte \& Tipton 2006, hereafter DT06; Chiravale 2006), and it reproduces three primary fluid instabilities: the Rayleigh-Taylor (RT) instability, which arises when a low density fluid supports a high density fluid under the influence of gravity or acceleration, the Richtmyer-Meshkov (RM) instability, which occurs when a shock interacts with a fluid of different acoustic impedance such as a density gradient, and the Kelvin-Helmholtz instability, which arises from velocity shear between two fluids,  even when they have otherwise identical properties.

In Scannapieco \& Br\"uggen (2008; hereafter SB08), the authors used the original DT06 model to study AGN-driven turbulence in galaxy clusters focusing purely on RT and RM driven turbulence. For our expanded model, we have added the additional contribution from  the KH instability, which is crucial to the problem we are studying.
	The DT06 model is based on the Navier-Stokes equations expanded to include a turbulent viscosity $\mu_{T}$ and pressure $P_T$ which are dependent on the eddy size $L$ and the turbulent kinetic energy $K$. To compute these properties we divide the flow into two components; writing velocity, for example, as the sum of mean $\tilde{u}$ and fluctuating $u''$ components:
	\begin{equation}
	u \equiv \tilde{u} + u'',
\end{equation}
where $\tilde{u}$ is the mass averaged variable $\tilde{u} \equiv \overline{\rho u}/\bar{\rho},$ $\overline{\rho u''} = 0,$  $\rho$ is the mass density, and the overbar represents an ensemble average over many realizations of the flow. This corresponds to an expansion about the mean flow and to first order yields the following evolutionary equations in 3D,
\begin{eqnarray}
	\frac{D \bar{\rho}}{Dt} &=& -\frac{\partial{\bar{\rho}\widetilde{u_{i}}}}{\partial{x_i}}, \label{masseqn}  \\ \nonumber \\
	\frac{D \rhobar F_r}{Dt} &=& \frac{\partial}{\partial{x_i}} \frac{\mu_T}{N_F} \frac{\partial{F_r}}{\partial{x_i}}, \label{spec}  \\ \nonumber \\
	 \frac{D \rhobar \tilde{u_i}}{Dt} &=& - \frac{\partial{P}}{\partial{x_i}} - \frac{\partial{\tau_{ij}}}{\partial{x_j}}, \label{veleqn} \\ \nonumber \\
	 \frac{D \rhobar \varepsilon}{Dt} &=& \frac{\partial{}}{\partial{x_j}} \frac{\mu_T}{N_{\varepsilon}} \frac{\partial{\varepsilon}}{\partial{x_j}} - P \frac{\partial{\widetilde{u_{j}}}}{\partial{x_j}} -\tau_{ij} \frac{\partial{u_i}}{\partial{x_i}} , \label{eeqn}
\end{eqnarray}
where $\rhobar$ is the mean density, $F_r$ is the mass fraction of species $r$,  $\rhobar \widetilde{u_i}$ is the momentum in the $i^{th}$ direction, $P$ is the mean gas pressure, and $\varepsilon$ is the internal energy per unit mass which, unlike in DT06 and SB08, includes both the thermal and turbulent kinetic energy components.  As discussed in Scannapieo \& Br\"uggen (2010), this formulation allows us to follow the model into the highly supersonic regime in which most of the internal energy is turbulent rather than thermal.
Turbulence affects the mean flow through the turbulent stress tensor $\tau_{ij}$ and the turbulent viscosity $\mu_{T}$
which is scaled in the energy equation by $N_{\varepsilon}=1$ and in the mass fraction equation by $N_F=1$. 
Finally, the Lagrangian time derivative is defined as
\be
	\frac{D}{Dt} \equiv \frac{\partial}{\partial{t}} + \widetilde{u_j} \frac{\partial}{\partial{x_j}},
\ee
where there is an implied summation over all dimensions. 

	These equations depend on the evolution of the eddy scale $L$ and the turbulent kinetic energy $K.$ Equations that include the diffusion, production, and  compression of these quantities are 
\ba
	 \frac{D \rhobar L}{Dt} &=& \frac{\partial}{\partial{x_i}} \frac{\mu_{T}}{N_L} \frac{\partial}{\partial{x_i}}L + \rhobar V + C_{C}\rhobar L \frac{\partial{\widetilde{u_{i}}}}{\partial{x_i}} \label{Leqn} \\ \nonumber \\
	 \frac{D \rhobar K}{Dt} &=& \frac{\partial}{\partial{x_j}} \frac{\mu_{T}}{N_\varepsilon} \frac{\partial{K}}{\partial{x_j}} - \tau_{ij} \frac{\partial{\widetilde{u_i}}}{\partial{x_j}} + S_K. \label{Keqn}
\ea
	In eqn.\ (\ref{Leqn}) the first term represents the diffusion of the eddy length scale as scaled by the turbulent viscosity $\mu_{T}$ and scale factor $N_L=0.5$.  The second term is the primary production term and is proportional to $V \equiv \sqrt{2K}$ and independent of the flow. The third term is the growth of eddies due to the expansion and compression of the mean flow. Finally, $C_C=1/3$ is a constant in the model and is determined by mass conservation in eddies as they are compressed. In eqn.\ (\ref{Keqn}),  which parallels eq.\ (\ref{eeqn}), the first term is the diffusion of turbulent kinetic energy and is scaled by $\mu_{T}/N_\varepsilon$. The second term is the work associated with the turbulent stress which drives the KH instabilities. Finally, the third term is a source term that drives RT and RM instabilities. 
	
Note that we assume that N$_F$ $\approx$ N$_{\varepsilon}$.
Pan \& Scannapieco (2010) studied the efficiency of mixing over a large range of high Mach number turbulent flows.  By comparing the scalar (e.g. mass fraction) dissipation time scale to the time scale for the total kinetic energy loss, they find that this ratio does not deviate much from one, which validates this choice.

	The primary source term for RT and RM instabilities is $S_K,$ which is represented by and defined by DT06,
\be
	S_K = \rhobar V \left(-C_B A_i \frac{1}{\rho} \frac{ \partial{P}}{\partial{x_i}} - C_D \frac{V^2}{L}\right),
\ee
where the coefficients $C_B=0.84$ and $C_D=1.25$ are fit to turbulence experiments. Physically, turbulent entrainment is described by $C_B$ which reduces any density contrasts, and $C_D$ is a drag coefficient that describes the dissipation of turbulent energy when the average length scale is proportional to $L.$ Likewise, $V \equiv \sqrt{2 K}$ is the average turbulent velocity, P is the pressure, $\rho$ is the density
, and $A_i$ describes the Atwood number in the $i^{th}$-direction. This is determined by,

\be
	A_i = \frac{\rhobar_{+} - \rhobar_{-}}{\rhobar_{+} + \rhobar_{-}} + C_A \frac{L}{\rhobar + L |\partial{\rhobar}/{\partial{x_i}}|} \frac{\partial{\rhobar}}{\partial{x_i}},
\ee
where $C_A=2$ is a constant of the model, $\rhobar_{+}$ and $\rhobar_{-}$ are the densities on the front and rear boundaries of a cell in the $i^{th}$-direction. 

	Additionally, and unlike DT06 and SB08, we include the full Reynolds stress tensor, constructed from mean velocities:
\be
	\tau_{ij} =  C_P \delta_{\rm ij}\rhobar { K} - \mu_{T}  \ \tau_{\rm KH} \left( \frac{\svel_i}{\spos_j} + \frac{\svel_j}{\spos_i} - \frac{2}{3} \delta_{\rm ij} \frac{\svel_k}{\spos_k} \right),
\ee
where $\delta_{\rm ij}$ is the Kronecker delta function, $\mu_{T}$ is the turbulent viscosity, $\tau_{\rm KH}$ is a function of the local Mach number which is calibrated to produce the correct KH growth rate as discussed below, and $C_P$ is a constant. The first term is the isotropic turbulent pressure and has a trace of $2\rhobar K$ when $C_P = 2/3$. The second term is the deviatoric tensor which has a zero trace (note the implied summation over all dimensions) and is the primary source of shear instabilities. 

Finally, the turbulent viscosity is calculated as.
\be
	\mu_{T} = C_{\mu} \rhobar L \sqrt{2K} \label{mueqn},
\ee
where $C_{\mu}=1$ is a constant. 
Table \ref{ctable} summarizes all model coefficients, their values, and their effects.

\begin{table}
\caption{}
\begin{tabular}{lll}
\hline
Coefficient & Value & Effect \\
\hline
\hline
$N_F$ & 1.0  & Diffusion of Species \\
$N_{\varepsilon}$ & 1.0 & Diffusion of Internal Energy \\
$N_L$ & 0.5  & Diffusion of L \\
$C_C$ & 1/3 & Compression Effects \\
$C_B$ & 0.84  & Buoyancy-driven turbulence \\
$C_D$ & 1.25  & Drag term on K \\
$C_A$ &  2.0 & Atwood Number \\
$C_P$ & 2/3 & Turbulent Pressure \\
$C_{\mu}$ & 1.0 & Turbulent viscosity \\
$\tau_{\rm KH}$ & variable & KH growth scaling \\
\hline
\label{ctable}
\end{tabular}
\end{table}

The numerical implementation of this model is divided into five steps: 	
\begin{enumerate}
	\item  Update the velocities using the turbulent viscosity in the fourth-order PPM solver before the turbulence package is called during the hydrodynamic step. 	
	\item Calculate the Reynolds stress tensor and update the turbulent kinetic energy, $K,$ as in eqn.\ (\ref{Keqn}).
	\item Use the updated value for $K$ and actualize the diffusive mixing terms in eqns. (\ref{spec}), (\ref{eeqn}), (\ref{Leqn}), and (\ref{Keqn}).
	\item Compute the contributions from the source terms as:
	\begin{enumerate}
		\item Calculate  $V \equiv \sqrt{2K}$.
		\item Add the $\rhobar V$ term  to the $L$ equation and use a leapfrog approach to add the source term ($S_K/\rho V$) to the $V$ equation.
		\item Write $K$ back as $K=  V^2/2$ a`nd update the turbulent viscosity as in eqn.\ (\ref{mueqn}). 
	\end{enumerate}
	\item Enforce a minimum time-step from turbulence as $dt \le (\Delta^2/\mu_T)/6$, where $\Delta$ is the minimum between $dx,dy$ and $dz$ in a given cell and $\mu_T$ is the turbulent viscosity in that cell. The minimum per block is calculated where a block in FLASH is composed of $nx \times ny \times nz$ cells. Finally the global minimum is calculated across all blocks.
		
\end{enumerate}	
	
\subsection{Sub-Grid Turbulence Model Tests}

\subsubsection{Rayleigh-Taylor Shock Tube Test}

	To verify the implementation of our model, we reconstructed the RT problem as described in $\S$5 of DT06 (and $\S$3.1 of SB08). Initially a 1 cm region was filled with two $\gamma$ = $5/3$ fluids with constant density, $\rho_1 =  1.0 \ \ {\rm g \, cm^{-3}}$ in the region from $x$ = 0.0 to 0.5 cm and $\rho_2 = 0.9 \ \ {\rm g \, cm^{-3}}$ from$x$ = 0.5 to 1.0 cm. A gravitational field acted in the $x$ direction with a constant acceleration of $9.8 \times 10^8 \ \ \rm{cm \ s^{-2}}$ or $10^6$ times the Earth's gravity. The initial temperature profile was calculated so that both fluids were in hydrostatic equilibrium and so that at x = 0.5 cm the temperature of the lower density fluid was $T_2$ = 50 K and the temperature in the higher density fluid was $T_1$ = 45 K. Finally, to test the mass fraction diffusion, we initialized each side with different generic mass fractions with atomic masses equal to hydrogen.
			
	Despite being described as a 1-dimensional problem, to test our implementation in FLASH we set up a 2-dimensional 50 ``block" by 1 ``block" region. Each block represented  8 $\times$ 8 simulation cells. Each test was allowed to refine up to a $l_{ref} = 4$ based on density and pressure profiles. This led to an initial cell size of  $1.0/50/8/2^3 = 3.1 \times 10^{-4}$ cm at the interface and a minimum resolution of $2.5 \times 10^{-3}$ cm. 
	
	Although there is an explicit turbulent time step that must be enforced, this implementation works very well with the AMR hydrodynamic time step imposed in FLASH. Initially when the center is fully refined, the hydrodynamic time step is shorter than the one imposed by turbulence. As the turbulent viscosity grows the time steps become comparable, however, because of the diffusion associated with turbulence the density contrast is smoothed. This allows the AMR to derefine these areas, which in turn increases the turbulent time step. Finally the turbulent time step reaches an equilibrium of 1/6 of the hydrodynamic time step at the end of the simulation after the whole volume has reached the lowest refinement level. 
	
	As described in DT06, this problem has an analytic solution for the evolution of $K$ and $L,$
\begin{eqnarray}
	K(x,t) &\approx& K_0(t) \left(1-\frac{x^2}{h^2(t)} \right), \\ \nonumber \\
	L(x,t) &\approx& L_0(t) \sqrt{1-\frac{x^2}{h^2(t)}}, \\ \nonumber
\end{eqnarray}
where $h(t)$ is the scale length for the interpenetrating fluid and is defined as,
\be
	h(t) = \alpha_b A(0) g t^2,
\ee
here $\alpha_b = 0.06$, $A(0)$ is the initial Atwood number = 0.05, g is the gravitational acceleration and is set to $9.8\times 10^8$ $\rm{cm s^{-2}}$,  $t$ is time in seconds, $K_0(t) = (dh/dt)^2/2,$ and $L_0(t) =  h(t)^2/2$.  Initially both $K$ and $L$ were set to a small values throughout the simulation except near the interface where we set both $K$ and $L$ to the analytic values at time of $50$ $\mu s$. 
	
\begin{figure*} 
\begin{centering}
\centerline{\includegraphics[width=6in,height=4.5in]{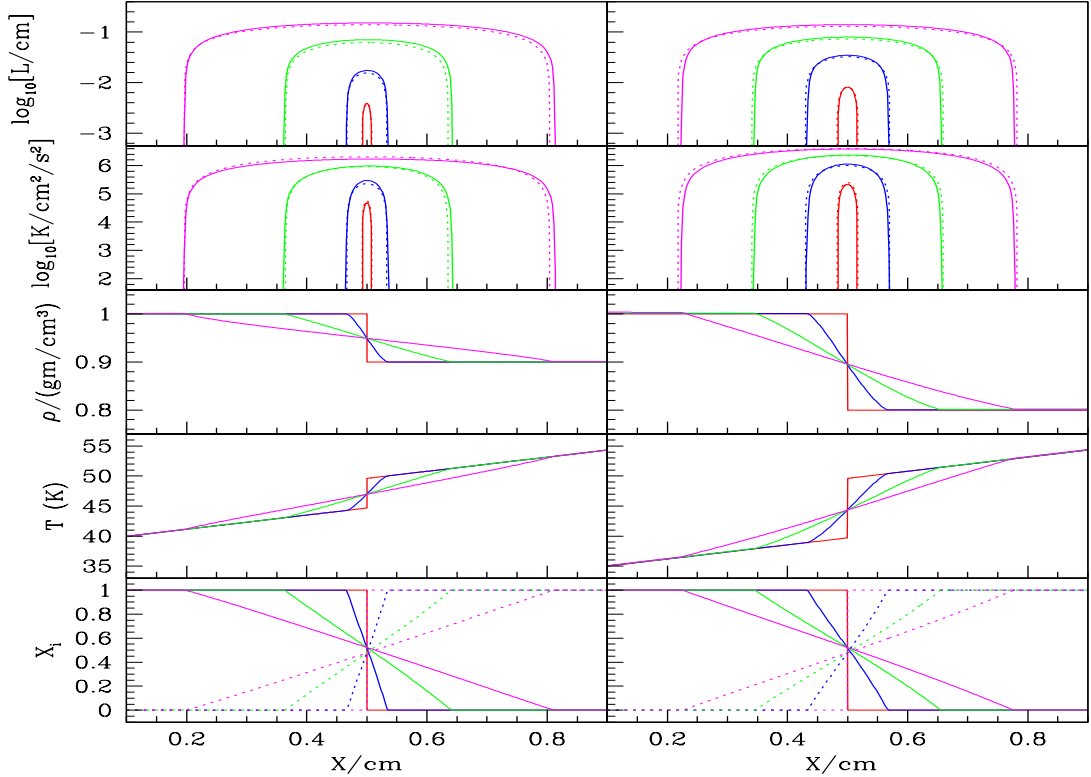}}
\caption{Evolution of the shock tube tests. The $\rho_2$ = 0.9 g cm$^{-3}$  case is given in the first column and the $\rho_2$ = 0.8 g cm$^{-3}$ case is given in the second column. {\it Top left panel}: Profiles of turbulent length scale at 50 (red), 100 (blue), 200 (green), and 300 (magenta) $\mu$s. In each case, the dotted lines are the analytic solution and the solid lines are the simulation results. {\it Second left panel}: Profiles of the kinetic turbulent energy  of the same case at the same times as the top panel. {\it Third left panel}: Density profiles at the same times. {\it Fourth left panel}: Temperature profiles. {\it Bottom left panel}: Profiles of species mass fractions. The dotted lines show the mass fraction of the species that was initially on the left side and the solid for the species on the right side. {\it Right panels:} same as the left except the profiles correspond to 50 (red), 100 (blue), 150 (green), and 200 (magenta) $\mu$s respectively. The $x$-axis and $y$-axis scales are the same in both columns.}
\label{RTC1}
\end{centering}
\end{figure*}

We find that our model matches the expected profiles for $K$ and $L$ at all times and as expected $L$ and $K$ evolve as $\propto$ $t^2$. As $K$ and $L$ quickly increase, the mixing layer also increases which promotes rapid mixing between the two fluids which can be seen in the species profiles.  We carried out the same test expect with a sharper density contrast  with $\rho_2$ = 0.8 gm/$\rm{cm^3}$ and $T_2$ = 40 K. The results are shows in the second column of Fig. \ref{RTC1}. Again, there is excellent agreement with the expected analytic profiles. 

\subsubsection{Shear Flow Test}

	To test the ability of our model to accurately model subsonic shear flow mixing we adapted the shear flow test problem described in \S 3 of  Chiravalle (2006). We began the problem with an initial velocity shear discontinuity at the origin. Left of the origin we set the $y$-velocity to $7.8 \times 10^4 \ \ {\rm cm s^{-1}}$ (M$_1$ = 0.46) while on the right we set the velocity to $1.09 \times 10^5 \ {\rm cm s^{-1}}$ (M$_2$ = 0.66). Pressure and density were held constant across the full domain at $1.72 \times 10^{10}$ erg cm$^{-3}$ and 1.0 g cm$^{-3}$ respectively. To study mass fraction diffusion, we again initialized each side with different mass scalars with identical properties. 
	
	In this case, the shear layer is expected to grow linearly with time as
\be
	\delta = 0.181 \, \Delta v \, t,
\ee
where $\delta$ is the width of the mixing layer, $\Delta v$ is the difference in velocity across the interface, and $t $ is the time (Chiravalle 2006).      
Unlike the test in Chiravalle (2006), we also added a small initial shear layer of size $\delta_{\rm init} =  0.1$ cm split equally through the interface with  $K  = 0.02 \ (\Delta v)^2$ and $L  = 0.2 \ \delta_{\rm init}$. With this setup, we were able to vary $\tau_{\rm KH}$ as a free parameter to approximate the expected growth rate. After 150 $\mu$s, the expected width of the shear layer is 0.923 cm, we find that $\tau_{\rm KH} = 0.20$ reproduces this result, as shown as the top line in Fig.\ \ref{LGfig}. This value is close to the one suggested by DT06 ($\tau_{\rm KH} \approx 0.1$). 

\subsubsection{Supersonic Shear Test}

	To extend our model into the supersonic regime, we compared our mixing layer widths to those obtained experimentally by Papamoschou $\&$ Roshko (1988), who measured the growth rate of the shear mixing  layer as a function of Mach number by forcing two fluids across each other at different relative speeds.  Defining the convective Mach number as
\be
	M_{\rm cl} \equiv \frac{|U_1 - U_2|}{a_1+a_2},
\ee
where $a_1$ and $a_2$ are the sound speeds and $U_1$ and $U_2$ are the fluid velocities in region 1 and 2 respectively,
Papamoschou $\&$ Roshko (1988) found that as $M_{\rm cl}$ increases, the mixing layer quickly decays asymptotically to 20\% of the subsonic mixing layer width.
To match this behavior at high Mach numbers we allow our variable $\tau_{\rm KH}$ to vary depending on a `local' Mach number which we define as
\be
	M_{\rm l} \equiv \frac{|({\bf \nabla \times {\tilde u}})| L}{c_s},
\ee 
where $|{\bf \nabla \times {\tilde u}}|$ is the magnitude of the curl of the mean velocity field, $L$ is the local eddy scale, and $c_s$ is the local sound speed. This local Mach number approximates $M_{\rm cl}$ and we use it to scale $\tau_{\rm KH}$ as
\be
\tau_{\rm KH} = \begin{cases} 
0.2 & M_{\rm l} \leq 0.3, \cr 
0.2 - 0.65 (M_{\rm l}-0.3) & 0.3 \leq M_{\rm l} \leq 0.6, \cr
0.00575  &  0.6 \leq M_{\rm l}.
 \label{ml2}
 \end{cases}
\ee
	    
To compare the result of this approach to the experimental measurements of $\delta$ as a function of time, 
we adapted the subsonic shear test from $\S$3.2 by changing the velocity on one side of the interface to match the the expected M$_{\rm cl}$. As above, we also initialized a small shear layer of size $\delta_{\rm init} =  0.1$ cm split equally through the interface with $K = 0.02 \ (\Delta v)^2$ and $L  = 0.2 \ \delta_{\rm init}$. $K$ and $L$ are initialized to zero everywhere else. Finally, by changing the initial $\Delta v$ (and thus M$_{\rm cl}$) and evolving for the appropriate time we measure the width of the final mixing zone.  

\begin{figure}
\begin{centering}
\includegraphics[scale=0.45]{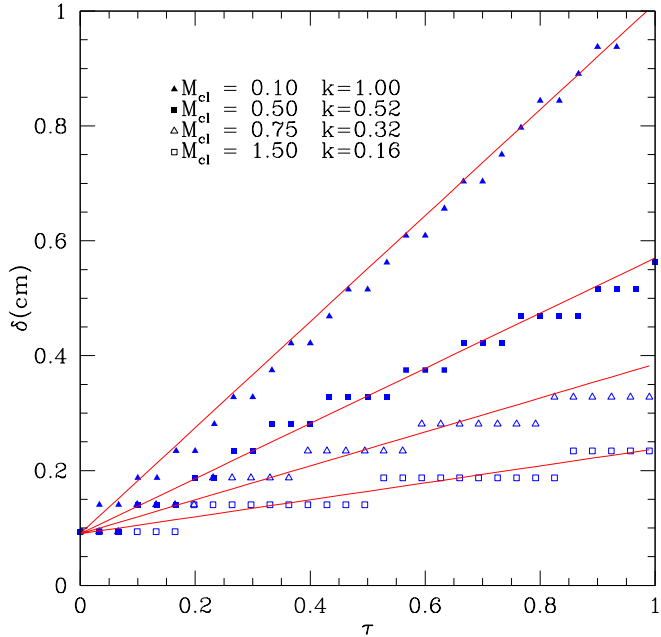}
\caption{Expected growth rate at different Mach numbers. 
The $y$-axis is the width of the mixing layer in cm and the $x$-axis is the normalized evolution time (the simulation time divided by the total evolution time $0.812 {\rm cm}/\Delta v,$).
 The red lines are the expect mixing widths from the $k$ values given in Papamoscho \& Roshko (1988)  while the blue points are the measured widths from our model.  Table \ref{c2table} summarizes the parameters used for the fits and for the model. The range of Mach numbers studied here is much larger than the range of Mach numbers found in our simulations, which vary between 0.3-0.7.}
\label{LGfig}
\end{centering}
\end{figure}

Fig.\ \ref{LGfig} shows the results of selecting a variety of values for M$_{\rm cl}$ and varying $\tau_{\rm KH}$.  Each set of points in this figure gives the mixing width measured as  the distance between the two points which correspond to  1$\%$ and 99$\%$ of the velocity difference.  Note that this is discretized due to this definition and the spatially discrete (AMR) structure of FLASH.  Also in Fig.\ \ref{LGfig} rhe solid lines are the theoretical mixing widths of the form
\begin{equation}
\delta = \delta_0 + 0.181 \, \Delta v  \, t \, k,
\label{deltaeqn}
\end{equation}
were $\delta_0$ is the initial mixing width, $\Delta v$ is the velocity difference between the two fluids, $t$ is time, and $k$ is a constant between 0 and 1. 

 In this figure the simulation time has been normalized  by the evolution time, defined as $0.812 \,{\rm cm}/\Delta v,$ the time required for each case to reach a final mixing width of $\delta$ = 0.923 cm if $k$ were equal to 1.  If $\tau_{\rm KH}$ is scaled correctly, then when fitting the mixing width using eqn.\ (\ref{deltaeqn}) $k$ will equal the expected mixing layer widths divided by the expected width as a function of Mach numbers as given by Papamoschou \& Roshko (1988) in their Fig.\ 16. As Fig.\ \ref{LGfig} shows, our model reproduces the expect linear growth extremely well across a range of Mach numbers. This range is much larger that seen our simulations, which have typical Mach numbers between 0.3-0.7. Table \ref{c2table} summarizes the initial setup parameters for the results in Fig.\ \ref{LGfig} as well as the final mixing widths and values for $k$. 

The growth rate of a shear layer is dependent primarily on the velocity and density ratio on either side of the shear discontinuity. This dependence has been studied by numerous authors (e.g. Brown \& Roshko (1974), Slessor \etal (2000) and references within) and as shown by Soteriou \& Ghoniem (1995) is small and for a given velocity ratio does not alter the growth rate very much.

\begin{table}
\caption{Parameters used for the Fig. \ref{LGfig}. The first 2 columns are the initial velocities on either side of the interface in units of Mach numbers, the 3rd column is the convective Mach number, the 4th column is the evolution time for each model, the 5th column is the final mixing width, and the 6th column is the parameter $k$ as defined in eqn.\ \ref{deltaeqn}.}
\begin{tabular}{llllll}
\hline
M$_1$ & M$_2$ & M$_{\rm cl}$($\frac{\Delta M}{2}$) & $\tau$ ($\mu$s) & $\delta$(cm) & k \\
\hline
\hline
0.46 & 0.66 & 0.10 & 150 & 0.923  & 1.0 \\
0.46 & 1.46 & 0.50 & 30 & 0.48 & 0.52 \\
0.46 & 1.96 & 0.75 & 20 & 0.32 & 0.32 \\
0.46 & 3.46 & 1.50 & 10 & 0.15 & 0.16 \\
\hline
\label{c2table}
\end{tabular}
\end{table}

\subsection{Radiative Cooling}

	Above $10^4$ K the cooling function is primarily controlled by atomic radiation and, at very high temperatures, bremsstrahlung radiation. These contributions are calculated from a table lookup using values calculated using CLOUDY (Ferland \etal 1998). Here we assume Case B recombination and consider only collisional ionization by use of the ``coronal equilibrium'' command of a metal free gas. Below 10$^{4}$ K the cooling is dominated by molecular line cooling and metal-line cooling. 
	
	Molecular cooling is described in Paper I, but now in addition, we have included metal line radiative cooling  in the optically thin limit. To simulate metals in our simulations, we define a generic mass scalar in FLASH, which is advected with any flows. For the cooling rates we use the tabulated results from Weirsma \etal (2008), which assume local thermodynamic equilibrium, and in all cases we use standard solar abundance ratios. The radiative rates are defined over a large temperature range, from $10^2$ K through $10^9$ K. The specific cooling rate for a given temperature is found using a table lookup from a  data file and scaling by the local metallicity.
		
\section{Modeling Outflow-Minihalo Interactions}

	Having described the new physical processes, we return our attention to the model developed in Paper I.  Again we assume a $\Lambda$CDM cosmology with  $h$ = 0.7, $\Omega_0$ = 0.3, $\Omega_{\Lambda}$ = 0.7, and $\Omega_{b}$ = 0.045 (e.g., Spergel \etal 2007), where $h$ is the Hubble constant in units of 100 km s$^{-1}$ Mpc$^{-1}$, and  $\Omega_0$, $\Omega_{\Lambda}$, and $\Omega_b$ are the total matter, vacuum, and baryonic densities in units of the critical density. The critical density for our choice of $h$ is $\rho_{\rm crit}$ = 9.2 $\times$ 10$^{-30}$ g cm$^{-3}$. 
	
	We begin with a neutral primordial minihalo, which is  composed of 24\% helium and 76\% hydrogen with a total mass of both dark and baryonic matter of M$_c$ = 3.0 $\times$ 10$^{6}$ M$_{\sun}$.  The initial minihalo has a total radial density profile given by Navarro \etal (1997):
\be
	\rho(R) = \frac{\Omega_0 \rho_c}{cx(1+cx)^2} \frac{c^2}{3F(c)} \ \rm {g \  cm^{-3}},
\ee
where $c$ is the halo concentration factor, $x \equiv R/R_{\rm c}$, $R_{\rm c} (= 0.393) \ {\rm kpc}$  is the virial radius, $F(t)$ $\equiv {\rm ln}(1+t)$ - $\frac{t}{1+t}$, and $\rho_{\rm c} = 6.54 \times 10^{-25}$ g cm$^{-3}$ is the mean cluster density. The baryonic matter is taken to be in hydrostatic equilibrium and follows an isothermal radial profile with a virial temperature of T = 1650 K:
\be
	\rho_{\rm gas}(R) = \rho_0 e^{-\frac{(v^2_{\rm esc}(0) - v^2_{\rm esc}(R))}{v^2_c}} \ \rm {g \  cm^{-3}},
\ee 
where the escape velocity is $v^2_{\rm esc}(xR_{\rm vir}) = 2 v_c^2 [F(cx)+cx(1+cx)^{-1}][xF(c)]^{-1}$ and $\rho_0$ = 2.16 $\times$ 10$^{-23}$ g cm$^{-3}$.

	Gravity is treated using the multigrid Poisson solver for self gravity of the gas (Ricker 2008) as well an additional component of gravitational acceleration due to dark matter. The initial metallicity of the halo and surrounding gas is set to zero, and the initial values of $K$ and $L$ are set to 1\% of the total internal energy and one parsec respectively. All other parameters are left at their fiducial values including the background UV radiation field (J$_{21}$ = 0.0). 
	
	The outflow is approximated by the Sedov-Taylor blast wave solution. We assume that the minihalo is at a distance $R_{\rm s}$ = 3.6 kpc and that the shock has a velocity of $v_s$ = 225 km s$^{-1}.$ By the time the shock reaches the minihalo it will have entrained a total mass of M$_{\rm s,total}$  = 4.4 $\times$ 10$^{7}$ M$_{\odot}$ with an associated surface density of $\sigma_{\rm s}$ = 2.6 $\times$ 10$^{5}$ M$_{\odot}$ kpc$^{-2}$. The input energy for this outflow was derived from SNe from the host galaxy, and was taken to be $E$ = ($\epsilon E_{55}$) erg where $\epsilon$ is the wind efficiency which is obtained from the fraction of the SNe energy funneled into the outflow and is set at our fiducial value of $\epsilon$ = $0.3$ and $E_{55}$ is the total SNe energy in units of 10$^{55}$ erg. 
	
	We initialized the left boundary with the same initial properties as our previous models. As in the rest of the simulation domain, we set $K$ and $L$ to 1\% of the internal energy and  one parsec respectively. To determine the initial metallicity of the shock, we followed the analysis from Scannapieco \etal (2004). We infer that roughly 2$\msol$ worth of metals are produced per $10^{51}$ ergs of energy in both Type II and pair instability supernova from Population III stars (Woosley \& Weaver 1995; Heger \& Woosley 2002). If we assume that half of these metals escape from the host galaxy and are funneled into the outflow, we can expect a total mass of metals of $M_{\rm metal} = 10^4 $ $E_{55}$ M$_{\sun}$. This leads to a metal fraction of the shock of $X_{\rm metal} = M_{\rm metal} /M_{\rm s,total} = 0.12$ Z$_{\sun}$, which we use as our fiducial value.

\section{Results}

	Our simulations were carried out in a rectangular box with an effective volume of 3.2 $\times$ 10$^9$ pc$^3$. The $y$-axis and $z$-axis  were both 1170 pc and ranged from (-585,585) pc. The $x$-axis was twice as long, 2340 pc, and ranged from (-585,1170) pc. The minihalo was centered at (0,0,0) pc and the shock originated from the left boundary with a velocity along the positive $x$ direction.  Both density and pressure were used as the refinement/derefinement variables, and we also forced derefinement after 7 Myrs in regions with density less than 3.26 $\times$ 10$^{-26}$ g cm$^{-3}$ outside a central sphere of radius 324 pc centered at (0,0,0) kpc. This had the advantage of derefining unimportant blocks, which was  especially important in the runs without subgrid turbulence, as turbulent mixing tends to smooth the density gradients, allowing the AMR to naturally derefine. 
	
A summary of the runs performed are given in Table \ref{truns}. We label them by whether they were high, medium, or low resolution (H, M, or L) and whether they used the subgrid turbulence package (WT or NT). In our new simulations metal cooling was always included, and to asses the impact of this cooling we  also include the fiducial run from Paper I  (HBN), noted by the asterisk, which is used to compare with run HNT. 
	
\begin{table}
\caption{Summary of the simulations. }
\begin{tabular}{lllll}
\hline
Name & $l_{\rm ref}$ & Resolution (pc) & Turbulence & Metal Cooling \\
\hline
\hline
LWT  & 4 & 18.22 & Y & Y \\
LNT   & 4 & 18.22 & N & Y \\
MWT & 5 & 9.11 & Y & Y \\
MNT  & 5 & 9.11 & N & Y \\
HWT  & 6 & 4.55 & Y & Y \\
HNT  &  6 & 4.55 & N & Y \\
HBN* & 6 & 4.55 & N & N \\
\hline
\label{truns}
\end{tabular}
\end{table}

\subsection{Hydrodynamic Evolution}

\begin{figure*}
\begin{centering}
\includegraphics[width=17cm, trim = 20mm  0mm 0mm 0mm]{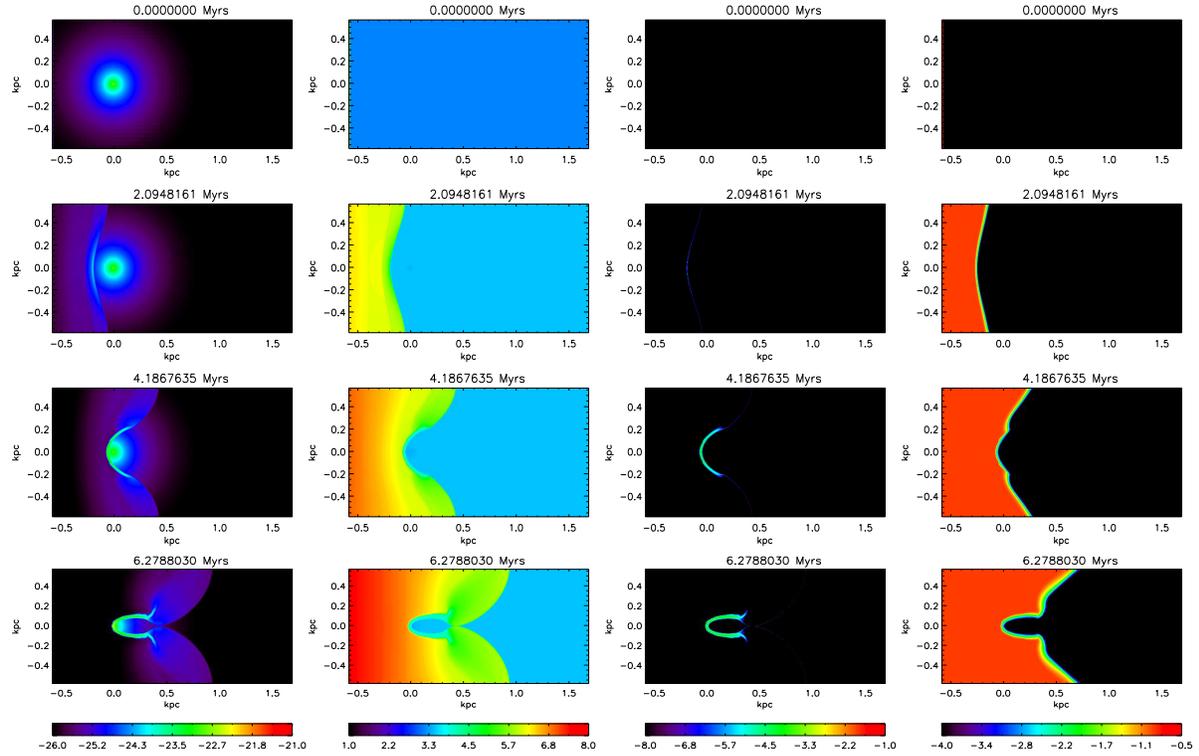} 
\caption{Evolution of run HWT from $t=0$ to the time at which the shock completely surrounds the cloud. Each image shows an $x$-$y$ slice through the center ($z$=0) of our simulation volume.   The first column shows logarithmic density contours from 10$^{-26}$ g cm$^{-3}$ to 10$^{-21}$ g cm$^{-3}$, which correspond to number densities between n $\approx$ 10$^{-2}$ cm$^{-3}$  and 10$^{2}$ cm$^{-3}$. The second column shows the logarithmic temperature contours from 10 K to 10$^{8}$ K, the third column shows the logarithmic H$_2$ mass fraction contours between X$_{\rm H_2}$ = 10$^{-8}$ to 10$^{-1}$, and finally the fourth column shows the logarithmic metal mass fraction contours  between Z = 10$^{-4}$ Z$_{\sun}$ to 10$^{-0.5}$ Z$_{\sun}$. }
\label{fig:evo1}
\end{centering}
\end{figure*}

\begin{figure*}
\begin{centering}
\includegraphics[ scale=.25, clip, trim = 20mm  0mm 0mm 0mm]{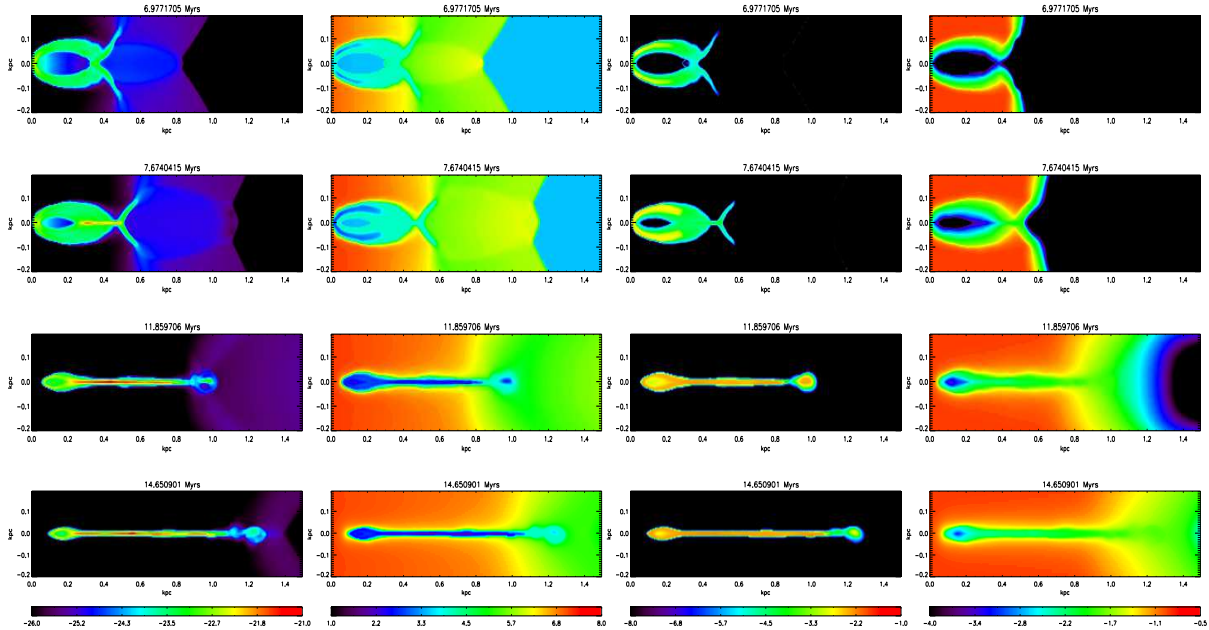} 
\caption{Evolution of run HWT from the time at which the shocks meet at the back of the cloud ($t$ = 6.797 Myrs), to the time at which the reverse
shock passes through the cloud ($t$ = 7.67 Myrs), to the collapse of the cloud ($t$ = 11.9 Myrs), and the end of the simulation ($t$ = 14.65 Myrs). Columns and rows are the same as in Fig.\ \ref{fig:evo1}. For this figure we have cropped the individual images along the $x$ and $y$ axes for clarity. Until $t$= 7.67 Myrs the molecule and metal distributions closely follow each other, but at later times molecule formation is  enhanced near the core of the cloud due to the reverse shock, which does not carry metals.}
\label{fig:evo2}
\end{centering}
\end{figure*}

	Figures \ref{fig:evo1} and \ref{fig:evo2} show the evolution of the minihalo from initial setup through the shock interaction and to the final collapse of the cloud, focusing on several important stages of evolution throughout these figures. The first row in Fig.\ \ref{fig:evo1} shows the initial  minihalo. Because the gas consists of neutral hydrogen and helium, it is unable to cool on its own.  Instead it remains in hydrostatic balance with a free fall time of
\be
t_{\rm ff} = \sqrt{\frac{3 \pi}{32 G \rho}} \approx 100 \ \rm{ Myrs},
\ee
that is roughly equal to the sound crossing time.

As the shock interacts with the minihalo, it begins to heat and ionize  the neutral  gas. The ionized gas then recombines and starts to catalyze the formation of H$_2$ and HD.  This results in a `hollow' distribution of molecules that is concentrated in the interacting regions at the front and sides of the cloud. The metal distribution closely follows this molecular distribution, because the shock is able to move faster in the less dense portions of the cloud. The shock fully envelopes the minihalo in a `cloud-crossing' time defined by Klein \etal (1994) as
\be
t_{\rm cc} = \frac{2 R_c}{v_s} \approx4.6 \ \rm{Myrs},
\ee
which occurs at about $t \approx$ 6.5 Myrs after the beginning of our simulations and is shown in the last row of Fig. \ref{fig:evo1}. 

The  time scale for H$_2$ formation is given by Glover \etal (2008) as
\be
t_{\rm H_2} = \frac{X_{\rm H_2}}{k_1 X_{\rm e} n},
\ee
where X$_{\rm H_2}$ and X$_{\rm e}$ are the mass fractions of H$_2$ and electrons respectively, $n$ is the number density, and $k_1$ is the reaction rate for the formation of H$^{-}$ a key reactant in formation of H$_2$. When the shock first interacts with the cloud, this time scale is very short since $n \approx$ 1 cm$^{-3}$ and the fraction of electrons (X$_e$) is relatively large.  However the cloud quickly reaches an abundance of $X_{\rm H_2} \approx 10^{-4}$ as shown in the third column of Fig.\ \ref{fig:evo1}. This fraction continues to grow as the cloud is surrounded, but does so at a much slower rate as $X_{\rm H_2}$ increases and X$_{\rm e}$ decreases. Again, the distributions of molecules and metals closely track each other.

At $\approx 7$ Myrs, after the shock meets on the back of the cloud, it creates a reverse shock, which begins to catalyze molecule formation, as shown  in the top row in Fig.\ \ref{fig:evo2}. It is here that the metal and molecule distribution diverge as the metals have not had the time to diffuse in the interior of the cloud. We define a turbulent mixing time scale as 
\be
t_{\rm mix} = \frac{d^2}{(\mu_T/\bar \rho)}  \ \rm{s},
\label{eqn:mix}
\ee 
where $d$ is the distance over which the metals are diffused, $\mu_T$ is the turbulent viscosity, and $\bar \rho$ the local density. By looking at the third and fourth columns of Fig.\ \ref{fig:evo2} and comparing the distributions of H$_2$ and metals at t = 7.67 Myrs, it is obvious that metals are deficient along the $x$-axis at $y$ $\sim$ 0.3 kpc.  If we approximate the distance that metals need to diffuse through as $d$ $\approx$ 10 pc and estimate the turbulent viscosity around the collapsing cloud to its post-shock value $\mu_T/\bar \rho \approx$ 25 pc$^2$ Myr$^{-1}$ then the mixing time scale is $\approx$ 4 Myrs.  Thus at $\approx 7+4=11$ Myrs, the distributions of molecules (shown by H$_2$) and metals once again follow each other, as can be seen in the third row of Fig.\ \ref{fig:evo2}.

The fourth row of Fig.\ \ref{fig:evo2} shows the final state of the cloud. Most of the material is  found in a small dense ribbon that is stretched along the $x$-axis and extends many times the initial virial radius away from the center of the dark matter halo.  This material is now much colder than it started with typical temperature around 100 K. The H$_2$ mass fraction abundance of this ribbon is around 10$^{-2.2},$ and it has a metallicity of about 10$^{-2}$ Z$_{\odot}$. 



\subsection{Model Dependencies}

\subsubsection{Effect of Turbulence}

\begin{figure*}
\centerline{\includegraphics[scale=0.85]{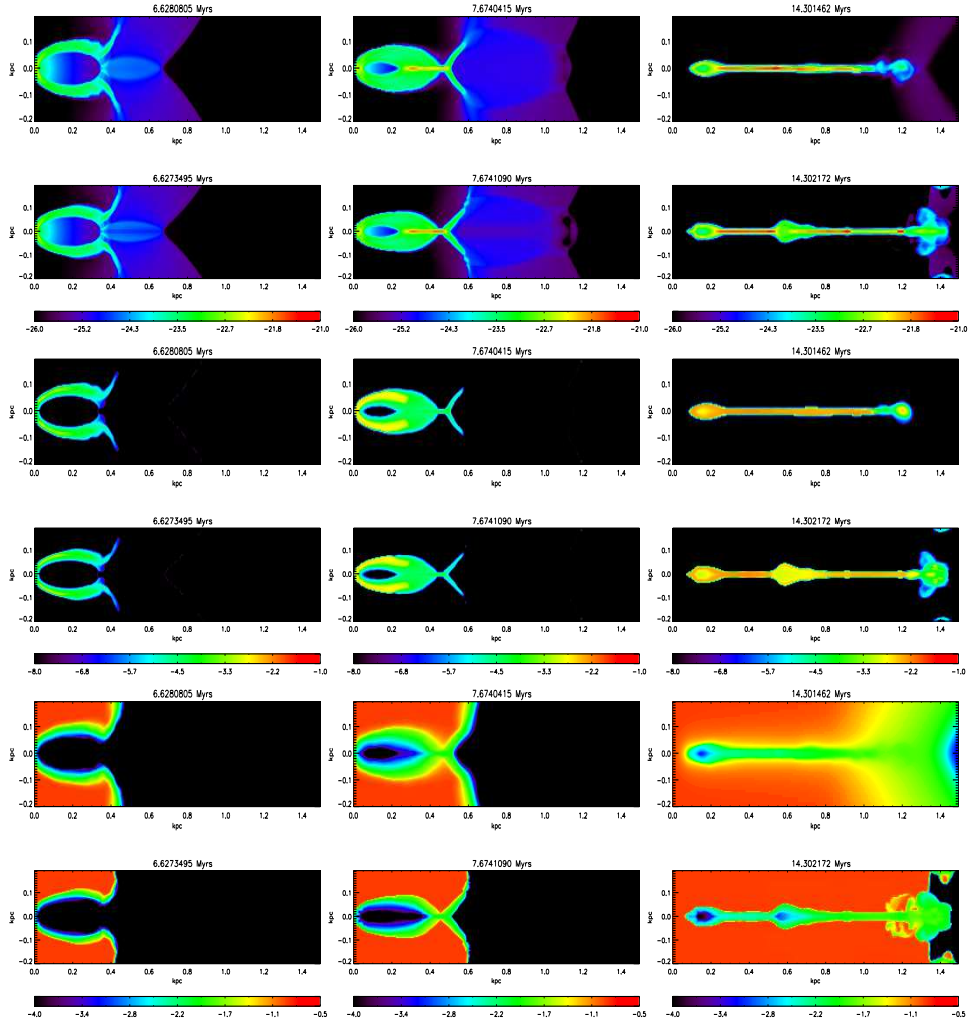}}
\caption{Late time comparison between a run with subgrid turbulence (HWT) and a run without it (HNT).  Each column represents a different snapshot in time. The top two rows shows logarithmic density contours from 10$^{-26}$ to 10$^{-21}$ g cm$^{-3}$. Rows 3 and 4 show the logarithmic contours of H$_2$ mass fraction from 10$^{-8}$ to 10$^{-1}$.  Finally, Rows 5-6 show the contours of metallicity in units of solar metallicity between 10$^{-4.0}$ to 10$^{-0.5}$ Z$_{\odot}$. In each set of rows the model with sub-grid turbulence (HWT) is on top of the model without it (HNT). Each image is a slice through the center of the domain along the $z$-axis. }
\label{fig:cturb}
\end{figure*}

Turbulence has two primary effects in our simulations: the diffusion of metals from the shock into the minihalo and the smoothing of sharp density contrasts.  Fig.\ \ref{fig:cturb} compares the difference between runs with (HWT) and without (HNT) our subgrid model for turbulence.
As expected many of the sharp density features found in the model without turbulence are diffused away in the model with subgrid turbulence. 
This is seen in the late-time panels in Rows 1 and 2 in Fig.\ \ref{fig:cturb} where there was a prominent lower density feature in HNT (at $x$ $\approx$ 0.65 kpc) that is not seen at all in HWT. Also absent are the smaller density features at the far end of the simulation domain along the $x$-axis. Although, in both cases the general result is the same: much of the mass has formed into a dense ribbon along the $x$-axis. 

The most striking difference between the two simulations is the metal distribution. HWT shows a very uniform metal abundance in the final cloud. Almost every portion has a final abundance of Z $\approx$ $10^{-2}$ Z$_{\odot}$ except  for a low density region near the initial center of the cloud. HNT however shows a much more uneven distribution with two regions of low metallicity; one at 0.15 kpc and the other at 0.6 kpc.   However, it is important to note that in both cases (HWT and HNT) the densest regions in both models have essentially the same final metal abundance, and thus the spread in metal distribution in the stars that are formed would be small with or without the inclusion of subgrid turbulence.

\subsubsection{Effect of Metal-Line Cooling}

\begin{figure*}
\begin{centering}
\includegraphics[scale=0.55]{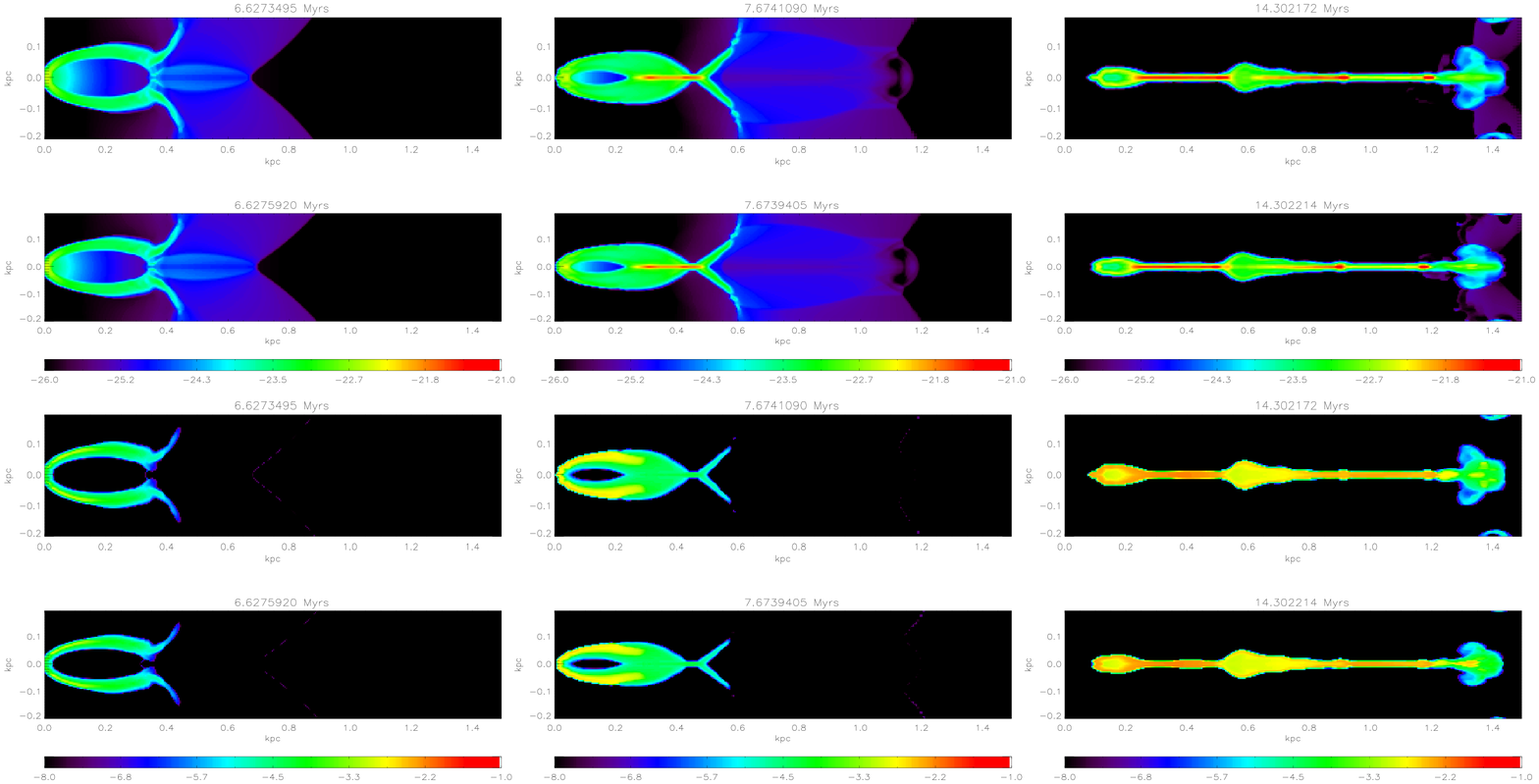}
\caption{Comparison between a run with metal cooling (HNT) and a run without it (HBN). Rows 1 and 2 show the logarithmic contours of density for HNT and HBN while Rows 3 and 4 show the logarithmic contours of H$_2$ mass fraction. Apart from the slight differences in the positions of similar features there is very little difference between the runs.}
\label{fig:cmetal}
\end{centering}
\end{figure*}

At temperatures below $T \approx 10^{4}$ K the primary coolants are molecules and low-energy metal-lines. Therefore the total cooling is  expected to be strongly dependent on differences between metal and molecule abundances. Figure \ref{fig:cmetal} shows the difference between a fiducial model with (HNT) and without (HBN) metal cooling, where HBN is taken from Paper I. There is little difference between these runs aside from slight changes in the positions of small structures. In both cases the same dense knots are found in essentially the same places. Furthermore, the abundances of molecular coolants are essentially identical and are not affected by the inclusion of metals. This suggests that metal cooling is not as important as molecular cooling, because otherwise the abundance of molecules in run HNT would be lower than that in run HBN due to the temperature dependence in the molecule formation rates. 

The time scale for H$_2$ cooling can be estimated as
\be
\tau_{\rm H_2} = \frac{1.5nkT}{n_e n_{\rm H_2} \Lambda(T)_{\rm H,H_2}},
\ee
where $n$ is the total number density, $k$ is Boltzmann's constant, $T$ is the temperature, $n_e$ and $n_{\rm H_2}$ are the number densities of electrons and H$_2$ respectively, and $\Lambda(T)_{\rm H,H_2}$ is the cooling rate as a function of temperature. Similarly for metal cooling the time scale is
\be
\tau_{\rm M} = \frac{1.5nkT}{n_e n_H \Lambda(T)_{\rm M} \frac{\rm{Z}}{\rm{Z_{\odot}}}} \ \rm{s},
\ee
where $n_e$ and $n_H$ are the number densities for electron and hydrogen and $\Lambda(T)_M$ is the cooling rate for metals. The ratio of these time scales is then
\be
\frac{\tau_{\rm H_2}}{\tau_M} = \frac{X_e}{X_{\rm H_2}} \frac{A_{\rm H_2}}{A_e} \frac{\Lambda(T)_{\rm M}}{\Lambda(T)_{\rm H,H_2}} \frac{\rm{Z}}{\rm{Z_{\odot}}},
\ee
where the number densities have been replaced using $X_i$ = $n_i A_i / \rho N_A$. To get an idea of the relative importance of each cooling method, we take  representative values for these variables at $t$ = 7.67 Myrs, an important point in the evolution of the cloud as it begins to collapse. We find that  $\tau_{\rm H_2} / \tau_{\rm M} \sim 10^{-2}$ which means that at this moment molecular line cooling is more important than metal line cooling. Overall, the inclusion of meal line cooling does not make a significant difference in the final state of the cloud. 

\subsubsection{Effect of Resolution}

\begin{figure*}
\begin{centering}
\includegraphics[scale=0.37]{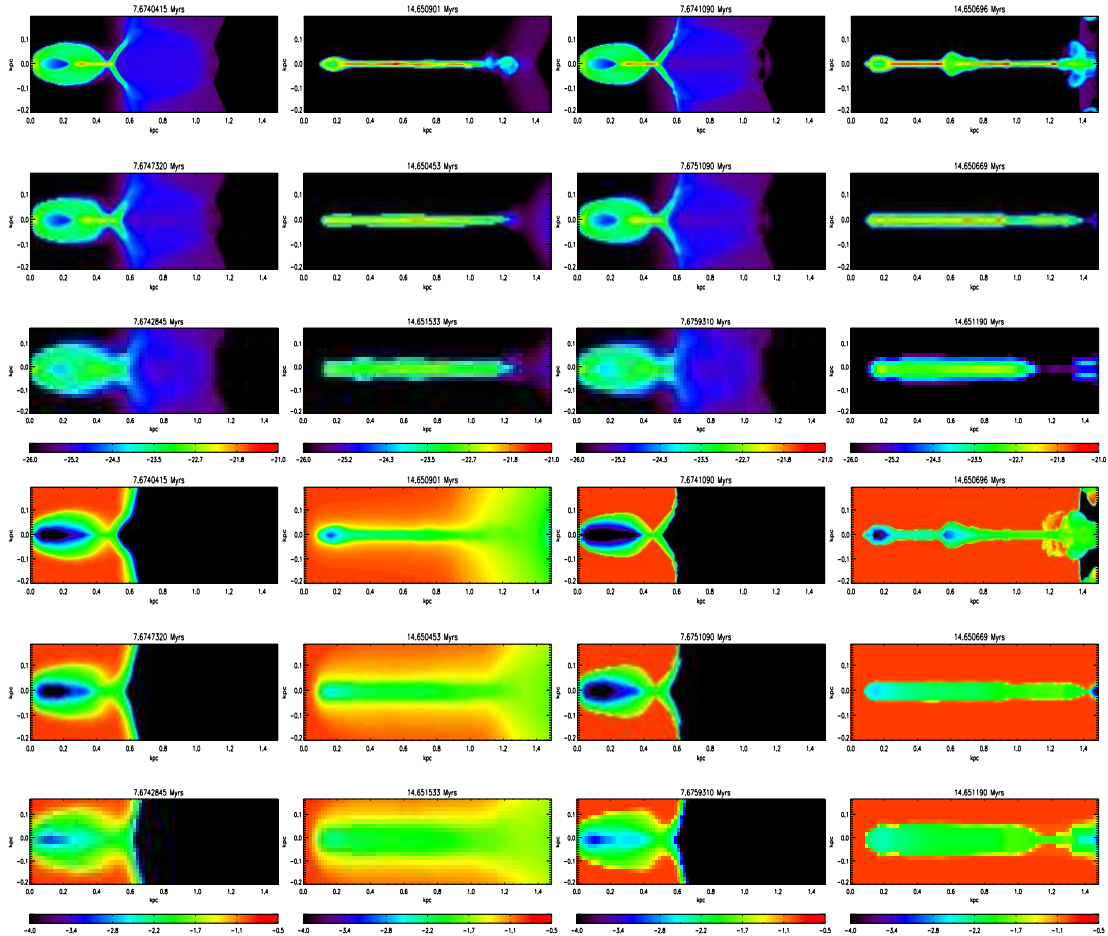}
\caption{The impact of maximum resolution on our runs with (left two columns) and without (right two columns) subgrid turbulence.  The top three rows show the logarithmic density contours at t = 7.67 Myrs and t = 14.65 Myrs and the bottom three rows show the logarithmic metallicity contours in units of solar metallicity. Rows 1 and 4 show the respective contours at the highest resolution with $l_{\rm max}$ = 6 (runs HWT and HNT), Rows 2 and 5 at $l_{\rm max} = 5$ (runs MWT and MNT), and Rows 3 and 6 at $l_{\rm max} = 4$ (runs LWT and LNT). } 

\label{fig:cres}
\end{centering}
\end{figure*}

Finally, we study the impact of resolution on our results. Fig.\ \ref{fig:cres} shows the result of our models over a range of maximum resolution levels. The left two columns show the results for models with turbulence while the right two columns show the results without turbulence. 
Generally, the outcome is independent of resolution. In the case of density the higher the resolution the smaller and more dense the final cloud becomes. However, the structure of the cloud is the same: it has been stretched into a ribbon and moved out of the dark matter halo. Furthermore, the metallicity distribution is almost indistinguishable between the runs with different resolution levels. For each choice of resolution level the final cloud is enriched to $\approx 10^{-2} Z/Z_\odot,$ both for runs with and without subgrid turbulence. 

Not shown is the difference in molecular abundances. Here there is a resolution dependence on the formation time scale for molecular coolants, as discussed in Paper I. However, the amount formed is always sufficient to cool the cloud enough to produce the same final outcome.  The final abundances are nearly identical at each refinement level except at the lowest resolution, which is different only outside of the dense regions in the final ribbon.
Aside from features that do not affect the final distribution of star-forming gas, our results are independent of resolution from $l_{\max} = 4$ to
$l_{\max} = 6.$

\subsection{Stellar Clusters}

\begin{figure}
\centerline{\includegraphics[scale=0.45]{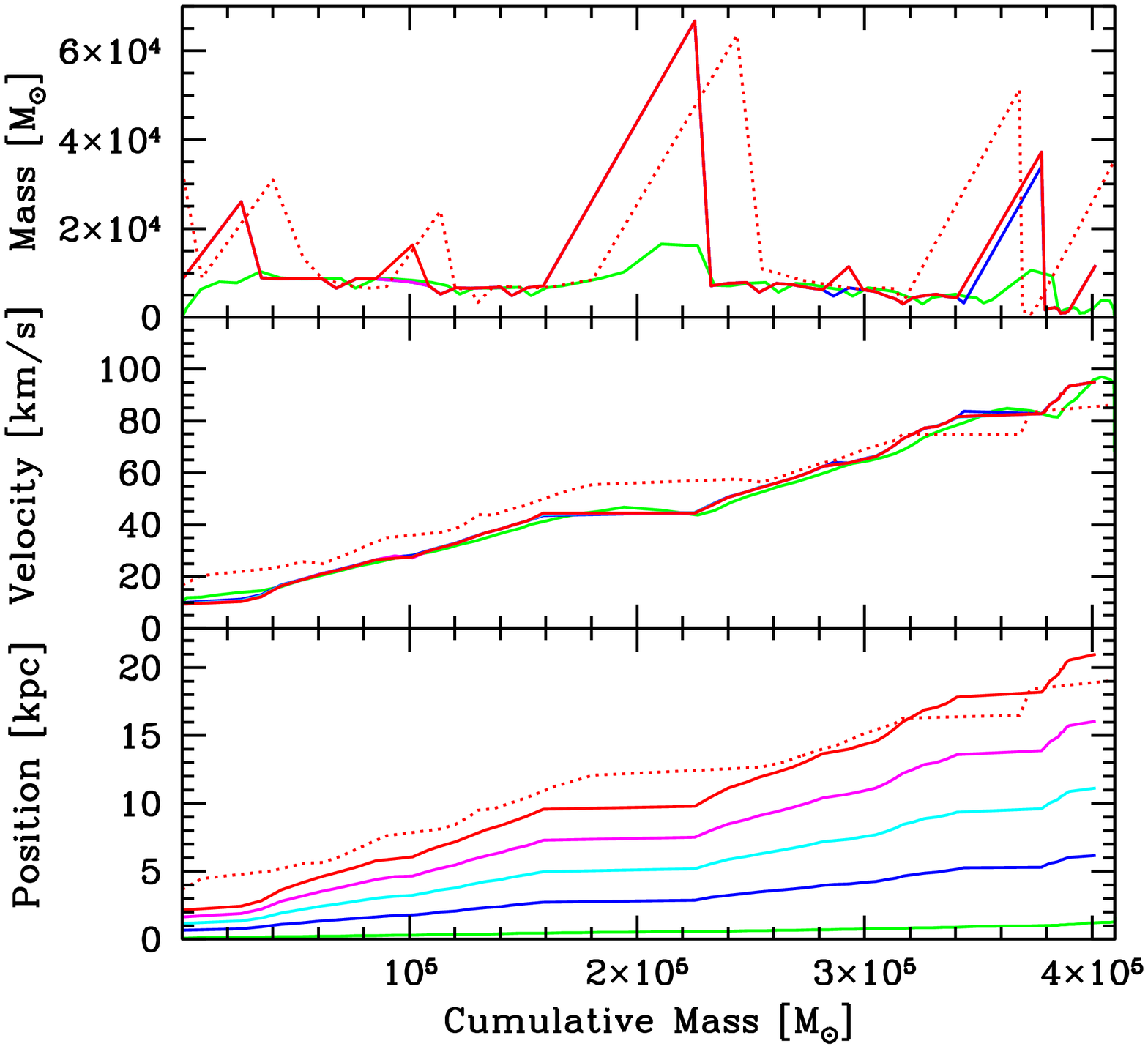}}
\caption{Long-term evolution of the distribution of particles after the end of the FLASH simulations. The $x$-axis gives the cumulative mass in units of  a  solar mass. The top panel shows the mass of each particle, the middle panel shows the velocity of each particle in units of kilometers per second, and the bottom panel shows the position of each particle in units of kpc. The solid lines are taken from run HWT, and the green line shows the initial profile at $t_f$ = 14.65 Myrs, the blue line shows the profile at 50 Myrs later, cyan shows the profile at $t_f$ + 100 Myrs, magenta shows the profile at $t_f$ + 150 Myrs, and the red line shows the $t_f$+200 Myrs. By $t_f$ + 50 Myrs most of the merging has finished and the largest clumps have formed. The dotted red line shows the $t_f$ + 200 Myrs distribution for run MWT and illustrates the difference between the same model at different resolutions. Although some minor differences are present, the same conclusions can be drawn from both runs. }
\label{fig:gmass}
\end{figure}

The final distribution of the dense clumps evolves over 100s of Myrs, a much longer time scale than the shock-minihalo interaction itself. To evolve our simulation over such a long time scale, we adopt the simple one-dimensional procedure described in Paper I.  Here we subdivide the $x$-axis into 100 evenly spaced bins from $x$ = 0 kpc to $x$ = 1.4 kpc. The mass for each bin is calculated by summing up the gas density from the simulation in a cylindrical volume with a radius of 24 pc and length of the bin. We also calculate the velocity of each new bin by summing together the momentum from each cell that goes into it and dividing by the total mass. 

This distribution is then evolved using a simple numerical model in which we assume all motion is along the $x$-axis and that pressure is not important at late times. The acceleration of each point is calculated by adding together the gravitational acceleration from all other particles as well as the gravitational acceleration from the dark matter halo. A leapfrog method is used to calculate the updated position and velocity from the calculated acceleration and updated velocity respectively. Finally, if a bin is evolved past the one in front of it, we merge these two by summing their mass and using conservation of momentum to compute its velocity. 

The model is evolved for 200 Myrs past the end of the simulation, and the results are shown in Fig.\ \ref{fig:gmass} for runs HWT and MWT.  As the gas continues to move outward the particles begin to attract each other and merge. By 50 Myrs after the FLASH simulation, most of the particles have merged and the motion of the remaining particles is purely ballistic. This can be seen in the top and middle panels of Fig. \ref{fig:gmass} as the lines begin to overlap each other. 

At the final time of 200 Myrs after the end of the simulation, we can identify one primary clump with a mass over 6.0$\times$10$^{4}$ M$_{\odot}$ with two smaller clumps with masses of 4.0$\times$10$^{4}$ M$_{\odot}$ and 2.5$\times$10$^{4}$ M$_{\odot}$ respectively. The two largest masses are far outside the dark matter halo, and are no longer bound to it. 

Comparing the solid (HWT) and dotted red line (MWT) we conclude that regardless of the resolution level the same general conclusions hold. There are some differences between the two resolution levels, primarily in the speed and position of the final clumps formed, but these are minor. In each case a nearly identical distribution of stellar clusters is formed:  one large cluster with a few smaller neighbors.

\subsection{Relation to Halo Globular Clusters}

As discussed in Paper I, shock-minihalo interactions could be a source of halo globular clusters, as the knots will continue to collapse into dense stellar clusters and the longest-lived of these stars will survive to the present day.  In fact the clusters in our simulations are very dense (n $\sim$ 10$^{2}$ cm$^{-3}$) and expected to become gravitationally bound to larger structures that form over cosmological time. 

There are several other properties of our simulated clusters that support this connection. Globular cluster masses are well defined by a Gaussian in log$_{10}$(M$_{\rm{GC}}$/M$_{\odot}$) with a mean mass of 10$^{5}$ M$_{\odot}$ and a dispersion of 0.5. This represents a spread in globular cluster masses of $3.0 \times$ 10$^{4}$ M$_{\odot}$ to $3.0 \times$ 10$^{5}$ M$_{\odot}$, which spans the range of the stellar cluster formed in our simulations. However the final cluster masses may depend on the initial mass in the minihalo a parameter study is required to determine this dependence. This is underway and will be presented in a future paper. 

 The lower mass limit for globular clusters seems to be set by several destruction mechanisms, which include mechanical evaporation (eg., Spitzer \& Thuan 1972) and shocking as the cluster passes through the host galaxy (eg., Ostriker \etal 1972). On the other hand, the upper mass limit seems to be a property of the initial population (eg., Fall \& Rees 1985; Peng \& Weisheit 1991; Elmegreen 2010). In fact, the upper limit of $\approx$ 10$^{6}$ M$_{\odot}$ roughly coincides with the virial temperature of T$\approx$10$^{4}$ K which corresponds to the limit at which atomic cooling becomes inefficient. 

The metallicity of halo globular clusters provides another constraint.  The intracluster metallicity distribution  is well described by a Gaussian with a mean value of $\left[\frac{Fe}{H}\right] \approx$ -1.6 and a dispersion of 0.3 (Zinn 1985; Ashman \& Bird 1993). The metallicity dispersion within a given globular cluster is small, usually within 0.1 dex (eg., Suntzeff 1993),  although in some cases additional late-time star-formation from reprocessed material may complicate the final observed distribution (\eg Piotto \etal 2007; D'Ercole \etal 2008; Bekki 2010).
Our model reproduces the expected mean abundance (Z $\sim$ 10$^{-2}$ Z$_{\odot}$) with an extremely homogeneous distribution given our initial abundance in the shock.

Note that our model keeps track of the velocity, $\sqrt{2 K},$ and eddy turnover scale, $L$, of bouyancy-driven and shear-driven turbulence, and assumes that below these lengths scales the flow will behave as fully developed turbulence.  In this case, as studed in detail in Pan \etal (2010), the mixing of metals is driven by a cascade process similar to that of the velocity field.  Using direct numerical simulations, Pan \etal (2010) showed that over a large range of Mach numbers which span the values in our in shock minihalo interactions, metals are mixed in on a time scale which is close to the time scale for energy decay, and that the dependence of this mixing time on the length scale at which pollutants are injected is also consistent with this cascade picture.

The final constraint comes from the study of tidally disrupted stars in globular clusters. Observations of globular clusters have shown tidal tails of stars actively being removed from these systems (Irwin \& Hatzidimitriou 1993; Grillmair \etal 1995) and have been used to argue that globular clusters do not  contain dark matter (Moore 1996; Conroy \etal 2010). If globular clusters reside within dark matter halos these tidal stars would be suppressed, which suggests that they formed via a different mechanism than galaxies. 

Our model naturally reproduces this feature as the shock moves almost all of the minihalo gas outside of the dark matter halo. As initially inactive gas interacts with the shock it is not only enriched by the metals to levels expected in globular clusters, the neutral gas is ionized and molecular species form via non-equilibrium chemical reactions. Not only does the gas have two methods of cooling below the T $\approx$ 10$^{4}$ K threshold, but the shock imparts enough momentum to remove it from the dark matter halo. The result is a population of stellar clusters with properties expected of halo globular clusters. 

Furthermore, the clusters formed in our simulations should be observable with the {\em James Webb Space Telescope (JWST)} as ribbons of dense star-forming gas that are elongated  versions of the more compact super-star clusters seen at low redshift (\eg Turner \etal 2000; Turner \& Beck 2004).  At $z$=10  {\em{JWST} } will have an angular resolution of $\approx $ 0.1" and our stretched ribbon of material will span $\approx$ 0.3", covering nearly three resolution elements. We can also estimate the apparent magnitude of an idealized cluster with a mass of 10$^{6}$ M$_{\odot}$. By assuming a single burst of star formation we calculate the apparent AB magnitudes in the restframe U, B, and V bands 5 Myrs after the burst of 32.0, 33.0, and 33.1 respectively (Bruzual \& Charlot 2003). Given the integration times appropriate for deep fields, these clusters can be detected by NIRCAM in the 4-6 $\mu m$ wavelength range (Gardner \etal 2006).  Although smaller clusters were formed for our fiducial model here, an upcoming parameter study will probe whether similar interactions with larger minihalos will form larger clusters with unique elongated morphologies.

\section{Summary}

	Cosmological minihalos provide the building blocks from which larger structures, such as galaxies, can form.  These structures require interactions from some outside source before they can evolve further, as they are filled with pristine, metal-free neutral atomic gas that surrounds the earliest galaxies.   Previous work has centered on using ionization fronts to induce star formation (Cen 2001), although follow up work using 3D hydrodynamic simulations showed that this would disrupt the cloud rather than induce molecule formation (Iliev \etal 2005; Shapiro \etal 2006).  On the other hand, galactic outflows provide a method of not only inducing molecule formation but also providing metals that can be mixed into the cloud, perhaps resulting in dense stellar clusters that evolve into today's halo globular cluster population.
	
	However, this scenario is only feasible if these metals were able to be efficiently mixed into the minihalo gas. While the metallicity difference between globular clusters can be quite large, in a given cluster the dispersion of (Fe/H) is quite small.  It is usually less than 0.1 dex, or a factor of $\approx$ 1.25 (Suntzeff \etal 1993) in the majority of clusters that do not show late-time star formation, which is likely from reprocessed material (\eg D'Ercole \etal 2008). We have shown here that in our fiducial interaction, most of the primordial gas is enriched to a near constant value of Z $\approx$ 10$^{-2}$ Z$_{\odot}$ over a time that roughly coincides with the H$_2$ formation time. 
	
	The chemical homogeneity is typically explained by a previous generation of supernovae, which either enriched the cloud from within, the `self-enrichment' case (e.g., Elmegreen \& Efremov 1997; Bromm \& Clarke 2002) or enriched the gas from which the cloud formed, the `pre-enrichment' case (e.g., Brown \etal 1995; Cen 2001; Nakasato \etal 2000; Beasley \etal 2003; Li \& Burstein 2003). Both of these scenarios have problems. In the `pre-enrichment' case it is unknown what this previous generation was, why it did not play a more significant role in the evolution of the cloud, and how it can enrich the cloud on such small time scales. However, the `self-enrichment' case is even more dire as it requires supernova exploding inside the cloud without completely unbinding the system. In fact, recent 2D simulations of such explosions have found that they do indeed destroy the minihalo over a wide range of masses and supernova energies (Whalen \etal 2008b).
	
Here we have provided a model that explains many of the properties of present-day halo globular clusters, through the interactions of a minihalo with a galaxy outflow. In our picture, the outflow performs three primary jobs. First, it provides the momentum to move the primordial minihalo gas out from its dark matter halo. Second, it provides a shock that catalyzes the formation of molecular species which serve as coolants at temperatures below 10$^{4}$ K. Finally, the outflow acts as a source of metals that are efficiently and uniformly mixed into the final stellar cluster. 
	
	We have shown that, when using the fiducial values for the  parameters described here, we create a distribution of high-redshift stellar clusters similar to a typical halo globular cluster observed today. Furthermore, the elongated nature of the forming cluster will be observable with the next generation of telescopes. In fact, {\em JWST} has the resolution to see the extended ribbon of clusters formed in our simulations and, given enough integration time, it has the sensitivity to image them. Specifically, the larger clusters (M $\approx$ 10$^{6}$ M$_{\odot}$) will have magnitudes (m$_{AB}$ $\sim$ 31) detectable by NIRCAM.   From a theory point of view, the next step will be to vary our model parameters over a wide range of physical  values, see what choices lead to realistic globular clusters, and determine how these can be constrained observationally. This study will be the focus of an upcoming paper.  

	We are grateful to Marcus Br\" uggen and Cody Raskin for useful conversations. We acknowledge the support from NASA theory grant NNX09AD106.  All simulations were conducted on the `Saguaro' cluster operated by the Fulton School of Engineering at Arizona State University. The results presented here were produced using the FLASH code, a product of the DOE ACS/Alliances-funded Center for Astrophysical Thermonuclear Flashes at the University of Chicago.

{}

\end{document}